# Origins of GaN(0 0 0 1) surface reconstructions

S. Vezian[a], F. Semond[a], J. Massies[a], D.W. Bullock[b], Z. Ding[b], and P.M. Thibado[b]

[a] Centre de Recherche sur l'Hetero-Epitaxie et ses Applications, Centre National de la Recherche Scientifique, Rue B. Gregory, Sophia Antipolis, F-06560 Valbonne, France

[b] Department of Physics, The University of Arkansas, Fayetteville, AR 72701, USA

**Abstract**
The reconstructions of the Ga polarity GaN(0 0 0 1) surface with and without trace amounts of arsenic and prepared by molecular beam epitaxy (MBE) have been studied with in situ reflection high-energy electron diffraction (RHEED) and scanning tunneling microscopy (STM). Various reconstructions are observed with RHEED by analyzing patterns while the substrate is exposed to a fixed NH3 flux or after depositing known amounts of Ga as a function of substrate temperature. In situ STM images reveal that only a few of these reconstructions yield long-range periodicity in real space. The controversial role of arsenic on Ga induced reconstructions was also investigated using two independent MBE chambers and X-ray photoelectron spectroscopy.

## 1. Introduction

Gallium nitride (GaN) is a technologically important member of the III–V compound semiconductor family of materials that are used to make high-power and optoelectronic devices and has been the subject of numerous studies [1–3]. Unlike Si-based devices which are primarily formed by ion implantation methods, III–V structures must be formed by epitaxial growth, depositing one plane of atoms on top of another until the entire device structure is formed. Naturally, surface structure plays an important role in the growth process. Furthermore, surface reconstructions have been shown to identify the polarity of the material [4]. Hence, there is both a technological and a basic science need to better understand III–V surface reconstructions. It is particularly true for the GaN(0 0 0 1) surface reconstructions which have been studied so far by a limited number of research groups [4–13]. There are two dominant techniques for preparing GaN by molecular beam epitaxy (MBE), growth using an ammonia source and growth using an atomic nitrogen source produced by a plasma cell [14–16]. Electron diffraction studies have reported numerous reconstructions with increasing substrate temperature and Ga coverage [5,6]. For these various phases, local structural information has also been reported, while several reconstructions have been examined theoretically and found to be stable [7]. Recently, a controversy over the role of trace amount of arsenic on the number and type of surface reconstructions has been reported [8–11]. Xue et al. have reported that GaN(0001)-(2×2) occurs due to excess Ga on the surface, while Ramachandran, et al. show this reconstruction happens when trace amount of arsenic is on the surface. In this study, we directly address the arsenic controversy by using two independent MBE system. One MBE system (Riber Compact 21) is new and has never been exposed to arsenic, while the other chamber is an older system that grew GaAs structures before.

## 2. Experimental

Experiments have been performed at CRHEA–CNRS [17]. Most of them were carried out in an ultrahigh vacuum (UHV) multi-chamber facility ($5 \times 10^{-11}$ Torr throughout) which contains a Riber solid-source MBE chamber. The growth chamber is equipped with a 25 keV RHEED system (Staib Instruments) and uses a valved-ammonia (NH$_3$) source as the nitrogen precursor. The growth chamber is connected to a surface analysis chamber with an X-ray photoelectron spectroscopy (XPS) system and to another UHV chamber containing an Omicron scanning tunneling microscope (STM). The samples were grown on Si(1 1 1) wafers (n-type, P-doped to have a resistivity between 0.025 and 0.05 Ω cm). Prior to the growth of a 1.5 μm thick GaN layer, a structure composed of successive epitaxial layers (AlN 40 nm/GaN 250 nm/AlN 250 nm) was grown at 920 C for AlN, 800 C for GaN, and using a NH3 beam equivalent pressure (BEP) of $5 \times 10^{-5}$ Torr. This growth procedure has been developed in order to overcome the specific strain problem encountered in the growth of GaN on silicon. More details can be found elsewhere [18]. For the RHEED measurements, surface reconstructions for a fixed NH3 flux were identified by either heating or cooling the substrate in 10 C increments, waiting 15 min, and recording the RHEED pattern. The symmetry of the surface at each temperature was then identified by analyzing the RHEED in two directions ([1 1 -2 0] and [1 -1 0 0]). In addition, the surface reconstruction transitions were measured without any NH$_3$ flux by heating the substrate in 10 C increments. Surface reconstructions for a fixed gallium (Ga) deposition were identified using a similar technique described above. The Ga was deposited using a deposition rate of 0.05 ML/s as determined from RHEED oscillations obtained previously during the growth of GaN with an identical Ga cell temperature. For the XPS and STM measurements, the GaN epitaxial layer was grown in the same manner as the RHEED sample, however the last 50 nm of the buffer layer was doped with Si ($10^{18}$ cm$^{-3}$) to facilitate STM imaging. Between the various STM studies, GaN was regrown on the substrate to produce a fresh surface. In order to prepare a particular surface reconstruction with as much long-range order as possible, the sample was annealed under the highest possible NH$_3$ flux and temperature that produced that reconstruction for as long as one hour. After this anneal, the NH$_3$ flux was ramped to zero at the same time the substrate temperature was ramped to the highest value that still produces the same reconstruction pattern. During the decrease in temperature the RHEED pattern was monitored to ensure that it remained unchanged. The sample was transferred to the XPS or STM without breaking UHV, and the data was collected at room temperature. The XPS measurements were acquired using an unmonochromatized Al Kα X-ray source (hν = 1486.6 eV). For the STM measurements, filled-state images were acquired using tips made from tungsten wire, with a sample bias of ±3–4 V and a tunneling current of 0.05–0.2 nA. All STM images have a (0001) plane subtracted from the data, but are otherwise unprocessed.

## 3. Results

Two XPS scans are shown in Fig. 1. The upper scan corresponds to a GaN(0 0 0 1)-(2×2) surface (obtained by exposing the surface to a Ga flux, see below), while the lower scan corresponds to a GaAs(0 0 1)-(2×4) surface. Notice that large Ga peaks are observed for each substrate. But, also notice that As peaks show up for the GaN scans. The arsenic is unintentionally present on the GaN surface and is assumed to be due to a background level of arsenic remaining in the MBE system from previous growths of GaAs. Unless specifically mentioned, for all future discussions of the GaN surface we assume that trace amounts of As are present on the surfaces with excess Ga. However, it should be noted that for the surfaces exposed to a large overpressure of NH$_3$ during the cooling procedure, corresponding for example to the preparation of the (2×2)-N (see below), the XPS measurements do not reveal the presence of As. The structural transitions between the different surface reconstructions as observed by RHEED for the GaN(0 0 0 1) surface as a function of NH$_3$ flux, Ga flux exposure and substrate temperature are shown in Fig. 2. At 800 C under normal N-rich growth conditions [19] the RHEED pattern observed during growth is mainly (1×1) with a faint diffuse trace of (2×2). We call this surface phase α(1×1)-N because of the N-rich growth conditions (see the lower panel of Fig. 2). Decreasing the substrate temperature while exposing the surface to only an NH3 flux of 50 μTorr BEP results in the formation of a clear (2×2)-N surface reconstruction which is characteristic of the (0 0 0 1) Ga polarity surface [4]. This transition occurs at about

550 C and the resulting pattern is very sharp and intense. At lower substrate temperatures (450 C) the RHEED pattern changes to (1×1). It has been shown that at this temperature and below, the cracking efficiency of ammonia becomes insignificant [20]. Therefore, we assign this surface phase as $\beta(1\times1)$-N. We speculate that this phase is linked to the non-dissociative adsorption of $NH_3$ molecules, while the high-temperature $\alpha(1\times1)$-N phase presumably results from dissociative chemisorption of $NH_3$ leading to the formation of $NH_2$ and/or NH chemisorbed radicals. Above 600 C a faint diffuse (2×2) reconstruction is still observed and we speculate that the $\alpha(1\times1)$-N corresponds to a disordered (2×2)-N phase. Increasing the substrate temperature reverses the reconstruction sequence at the same temperatures. The zero $NH_3$ flux data series, where the transitions only happen when the substrate temperature is increased from the $\beta(1\times1)$-N phase, is shown directly above the non-zero data. Actually, the $\beta(1\times1)$-N phase is unstable under UHV conditions (i.e. without impinging $NH_3$ flux) and exposing the surface to the electron beam of the RHEED gun immediately provokes the $\beta(1\times1)$-N to (2×2)-N transition.

To induce other reconstructions, Ga was deposited on the surface without NH3 present. Ga deposition at high temperatures (650–800 C) resulted in the RHEED pattern changing from (1×1) to (2×2). The half-order RHEED streak intensity changes as a function of the Ga exposure time is shown in Fig. 3(a). From this graph, the Ga coverage of the (2×2) reconstruction could be estimated at one monolayer. In this temperature range, excess Ga is probably desorbed, since the (2×2) reconstruction persist for many monolayer worth of Ga. During annealing at 800 C without Ga impinging flux, a transition from (2×2) to (1×1) is observed. On the other hand, starting from a (2×2)-Ga surface and increasing the Ga exposure at slightly lower temperatures (450–600 C) resulted in the reconstructions changing from (2×2) to (4×4). The intensity change in a quarter order streak is shown in Fig. 3(b). This graph is separated in two time segments. In the first time segment, the quarter fractional order RHEED intensity increases and is maximum at a Ga deposition of 0.6 ML. In the second time segment, corresponding to a Ga deposition increase from 0.6 to 1.0 ML, the quarter fractional order RHEED intensity decreases and the pattern becomes (1×1). This transition is reversible and we observed a transition from (1×1) to (4×4) to (2 · 2) while annealing at 600 C without impinging Ga. Note that the (4×4) is only obtained from Ga deposition on the (2×2)-Ga surface. Starting directly from a (1×1) and exposing the surface to the Ga flux at a temperature below 650 C does not change the RHEED pattern.

Characteristic STM images of the GaN(0 0 0 1) surface after preparing the (2×2)-N surface reconstruction are shown in Fig. 4. A typical large-scale STM image is shown in Fig. 4(a). Here each gray level represents a terrace which is separated from the next by a GaN monolayer high step (0.26 nm). The surface tends to favor long straight steps running along the [1 1 -2 0] direction on alternate terraces, while favoring triangular or jagged step edges for the other set of terraces [21,22], similar to other wurtzite systems like MnAs or ZnO [23,24]. A higher-magnification image is shown in Fig. 4(b), which shows again the alternately straight or jagged terrace edge. In addition, small triangular-shaped pits are now clearly visible within any given terrace. An even higher-magnification image of the surface is shown in Fig. 4(c). Even at this scale it is not possible to see the periodicity of the surface reconstruction. Nevertheless, the triangular shape of the pits persists down to the atomic scale. The orientation of the pits precisely follows the triangular step edge directions. These specific pits, which have one GaN monolayer depth, are delimited by an equilateral triangle and are not related to the presence of a screw type dislocation [21].

Characteristic STM images of the GaN(0 0 0 1) surface after preparing the Ga induced (2×2) surface reconstruction are shown in Fig. 5. A typical large-scale image is shown in Fig. 5(a). Again, each gray level represents a terrace that is separated from the next by a monolayer high step. Notice that this surface tends to favor steps that are more rounded in the (0 0 0 1) plane. A high magnification image is shown in Fig. 5(b), which shows large terraces with still a hint of the alternately jagged-smooth terrace edges. Notice that even though the terrace edges do get very close to each other, double-height steps never form. This surface does not have as many large triangular pits as the (2×2)-N. An even higher-magnification image is shown in Fig. 5(c). Here the origin of the (2×2) surface reconstruction is more clear. The diamond shaped box overlaid on Fig. 5(c) highlights the conventional unit cell. Notice, a fair fraction of the image appears ''fuzzy.'' Underneath the fuzzy areas, a well-ordered surface can sometimes be seen. This is typical of the Ga induced (2×2) surface reconstruction we have observed [17].

Characteristic STM images of the GaN(0 0 0 1) surface after preparing the (4×4) surface reconstruction by Ga exposure of the (2×2) at 500 C are shown in Fig. 6. A typical large scale STM image is shown in Fig. 6(a). This surface has even smoother terrace edges, but still maintains a hint of the alternately smooth and jagged terraces edges. A higher-magnification image is shown in Fig. 6(b), which shows a few terraces each of which is peppered with irregular shaped pits. At this magnification, it is difficult to see the different terrace edge types. An even higher-magnification image of the surface is shown in Fig. 6(c). This image reveals the origin of the (4×4) surface reconstruction. The box overlaid on Fig. 6(c) highlights the conventional diamond shaped unit cell. Other than the frequent occurrence of pit-type defects, this surface is well ordered and easily imaged with STM.

During the (4×4) to (2×2) transition obtained by annealing under UHV at 600 C, the sample was cooled down and imaged with the STM as shown in Fig. 7. Two typical large-scale STM images are shown in Fig. 7(a) and (b). These images were taken within a span of 45 min of each other and in the same location on the surface. Notice the island on top of the terrace has changed its shape between scans and became smaller. A higher-magnification image of the terrace is shown in Fig. 7(c), which shows a similar image to the (2×2) shown previously in Fig. 5(c). More interesting is a higher-magnification image of the island which is shown in Fig. 7(d). This image reveals a new reconstruction not observed with RHEED, but has the sqrt(7) × sqrt(7) R19.1 symmetry. As this region is imaged, the atoms are detaching themselves from the island and moving around on the terrace. This suggests the origin of the ''fuzzy'' sections of the (2×2)-Ga STM images is excess Ga roaming around the surface.

## 4. Discussion

It is insightful to compare and contrast the GaN(0 0 0 1) surface with GaAs surfaces, since they have been studied for a longer period and the material system is more mature [25,26]. The GaN(0 0 0 1) bulk-terminated surface consists of a plane of Ga atoms in a triangular lattice each with one broken bond, making this a fairly stable as well as polar surface. This is very different from the GaAs(0 0 1) bulk-terminated surface, which is a plane of Ga atoms in a square lattice each with two broken bonds, making this surface less stable. However, the GaN(0 0 0 1) bulk-terminated surface is identical to the GaAs(1 1 1)A surface (Ga-polarity). It should be noted that for the case of GaAs(1 1 1)A surface, only a (2×2) surface reconstruction has been observed [25]. From this we would expect a small number of surface reconstructions to form on the clean GaN(0 0 0 1) Ga polarity surface. This bring us to the discussion of the role of the As contamination clearly evidenced by XPS on surface reconstructions obtained by exposure to a Ga flux (Fig. 1). It is known that arsenic segregates to the GaN surface during growth and floats along the growth front [27,28]. By comparing the GaAs (2×4) XPS scan with the GaN (2×2) XPS scan, we estimate the coverage of arsenic to be 0.25 ± 0.05 MLs. This value is consistent with the previous estimate made by Ramachandran et al. for a (2×2) surface reconstruction obtained only by As exposure [8,11]. Even though the XPS shows trace amounts of As on the surface, this does not necessary mean that the (2×2) and (4×4) reconstructions we observed are due to its presence. To further test this, we prepared GaN in a separate and independent MBE chamber which has never been used before for the growth of GaAs and has only been loaded with Ga and $NH_3$ sources. Although it was easy to obtain the (2×2)-N under $NH_3$ exposure in this As-free MBE chamber, it was impossible to produce by Ga exposure the (2×2) and (4×4) reconstructions shown in Figs. 2 and 3. This provided further evidence that arsenic plays a crucial role in the stabilization of these Ga-rich surface reconstructions, as previously reported and theoretically predicted by Ramachandran et al. [8]. Note that, as also reported by Ramachandran et al., we have verified that a (2×2) reconstruction can be obtained from a (1×1) RHEED pattern by exposing the surface to a standard arsenic effusion cell (As$_4$ flux) in the 650–750 C temperature range and without NH$_3$ impinging flux. However, the (4×4) cannot be obtained by only direct exposure of the surface to an As$_4$ flux.

Considering now the difference between the (2×2) and (4×4) reconstructions obtained by Ga exposure in presence of As residual contamination, we found that the (4×4) surface reconstruction is produced only after adding more Ga to the (2×2). This is contrary to an earlier report [9,10] in which the (4×4) surface reconstruction was produced by annealing the (2×2) and was therefore associated with a lower Ga coverage. Atomic structural models have then been proposed which show the (4×4)-Ga surface reconstruction having less Ga than the (2×2)-Ga, which is inconsistent with our results [9,10]. It is possible that in these previous studies Ga droplets on the surface are supplying, over time, the needed extra Ga. One significant benefit of our study is that we use NH$_3$ versus a nitrogen plasma source, making the growth conditions N-rich versus Ga-rich, allowing better control of the amount of Ga actually on the surface [29].

Another puzzling reconstruction is the sqrt(7) × sqrt(7) R19.1, which was only observed in small regions of the sample and only with STM (not RHEED). Conveniently, this reconstruction was observed in parallel with the (2×2). From this we know the sqrt(7) × sqrt(7) R19.1 islands are on top of a (2×2) terrace and requires more Ga to form, which is consistent with earlier reports [9,10] and seems to indicate that this phase is in some way intermediate between the (4×4) and the (2×2) phases.

## 5. Conclusion

The reconstruction phases of the GaN(0 0 0 1) Ga polarity surface prepared by MBE using ammonia have been mapped out as a function of NH$_3$ flux, Ga exposure and substrate temperature with in situ RHEED. We have found two different (2×2) reconstructions depending on the experimental procedure. Using N-rich conditions at low temperatures we see a (2×2)-N reconstruction. We have shown that this reconstruction is characteristic of a clean GaN(0 0 0 1) Ga polarity surface exposed to ammonia. On the other hand, in a chamber previously used for GaAs growth, we found trace amount of arsenic on the surface of GaN when an excess of Ga exists. In addition, these samples formed (2×2) and (4×4) surface reconstructions. In a separate chamber never used before for GaAs growth, these GaN reconstructions could not be reproduced. We therefore conclude that these reconstructions are associated with an As surface contamination as was previously reported by Ramachandran et al. for the (2×2) reconstruction [11]. However, adding Ga on the (2×2) is necessary to produce the (4×4) which means that in addition to As contamination, Ga plays a key role in this reconstruction. Moreover, the associated Ga surface coverage is higher for the (4×4) than for the (2×2), contrary to earlier reports.

## Acknowledgements

The authors thank N. Grandjean and N. King for useful discussions. This work was supported in part by CRHEA–CNRS, and by the National Science Foundation of the United States of America under grant no. FRG-DMR-0102755. P.T. is grateful for the financial assistance provided by the Centre National de la Recherche Scientifique/Departement Sciences Physiques et Mathematiques (France).

Fig. 1. (upper line) XPS spectra taken from a GaN(0 0 0 1)-(2×2) surface obtained by Ga deposition (see text). This scan shows both Ga and As peaks. (lower line) XPS spectra taken from a GaAs(0 0 1)-(2×4) surface. This scan is shown for calibration and comparison purposes. Both scans are shown to scale and were collected over the same time period. Notice the smaller intensity of the As peaks for the GaN sample and the Mo peak due to the sample holder.

Fig. 2. RHEED derived surface reconstruction transition temperatures for GaN(0 0 0 1) as a function of incident NH$_3$ BEP and Ga deposition. The high BEP NH3 data is taken using the flux normally used during

GaN growth and are reversible. The zero NH$_3$ pressure transition temperatures are only for increasing temperature. The Ga deposition was determined by RHEED (see Fig. 3).

Fig. 3. Fractional RHEED intensities and associated RHEED pattern images taken during Ga exposure producing the transition between: (a) (1×1) to (2×2) at 650 C and (b) (2×2) to (4×4) to (1×1) at 500 C. The recorded intensities correspond respectively to a half-order (a) and a quarter-order (b) RHEED streak (the window corresponding to the intensity measurement is indicated on the RHEED images).

Fig. 4. STM images for the GaN(0 0 0 1)-(2×2)-N surface reconstruction: (a) 300 nm × 300 nm STM image showing flat terraces; (b) 90 nm × 90 nm STM image showing a close up of the alternately jagged and straight terrace edges; (c) 18 nm × 18 nm STM image showing the triangular shape of the pits which form on the terraces. The sample bias is 4 V and the tunneling current is 100 pA except for (c) which is 50 pA.

Fig. 5. STM images for the GaN(0 0 0 1)-(2×2) surface reconstruction prepared by Ga deposition: (a) 300 nm × 300 nm STM image showing flat terraces; (b) 90 nm × 90 nm STM image showing a close up of the alternately jagged and straight terrace edges; (c) 18 nm × 18 nm STM image showing the ''fuzzy'' atomic surface reconstruction. A (2×2) unit cell is drawn. The sample bias is -3 V and the tunneling current is 100 pA.

Fig. 6. STM images for the GaN(0 0 0 1)-(4×4) surface reconstruction prepared by Ga deposition: (a) 300 nm × 300 nm STM image showing flat terraces (sample bias = -3 V, tunneling current = 100 pA); (b) 90 nm × 90 nm STM image showing a close up of the terraces and a hint of the atomic surface reconstruction is visible on the middle terrace (sample bias = -3.5 V, tunneling current = 100 pA); (c) 18 nm × 18 nm STM image showing well-order atomic structure of the (4×4) surface reconstruction (sample bias = -3 V, tunneling current = 50 pA). A (4×4) unit cell is drawn.

Fig. 7. STM images for the mixed GaN(0 0 0 1)- sqrt(7) × sqrt(7) R19.1/(2×2) surface reconstruction: (a) 90 nm × 90 nm STM image showing an island with the sqrt(7) × sqrt(7) R19.1 reconstruction (sample bias = -3 V, tunneling current = 100 pA); (b) 90 nm × 90 nm STM image of the same location as (a) but 45 min later, notice the island shape and size has changed (sample bias = -3 V, tunneling current = 50 pA); (c) 25 nm × 25 nm STM image showing a zoomed in image of the terrace shown in figure (b). A (2×2) conventional unit cell is drawn. (d) 25 nm × 25 nm STM image showing a zoomed in image of the island shown in figure (b). A sqrt(7) × sqrt(7) R19.1 unit cell is drawn.

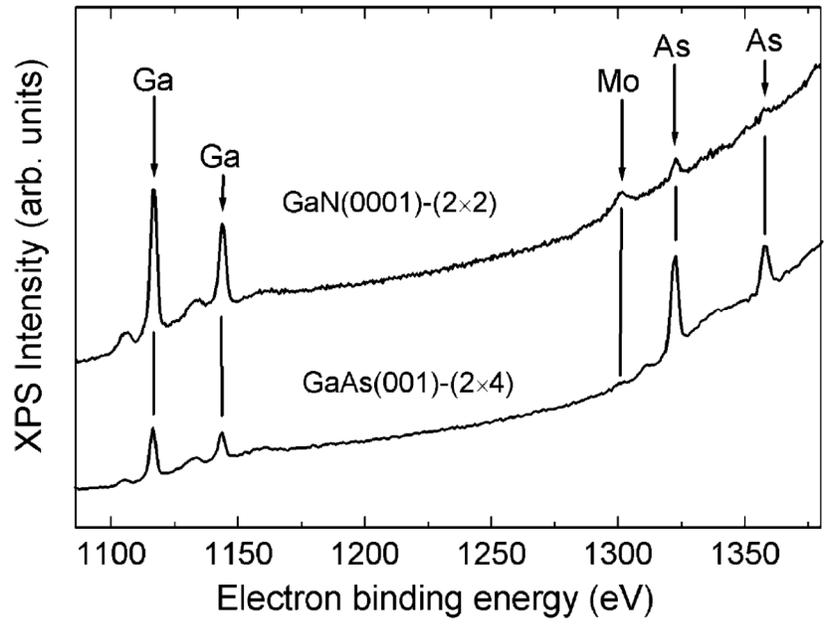

Fig. 1

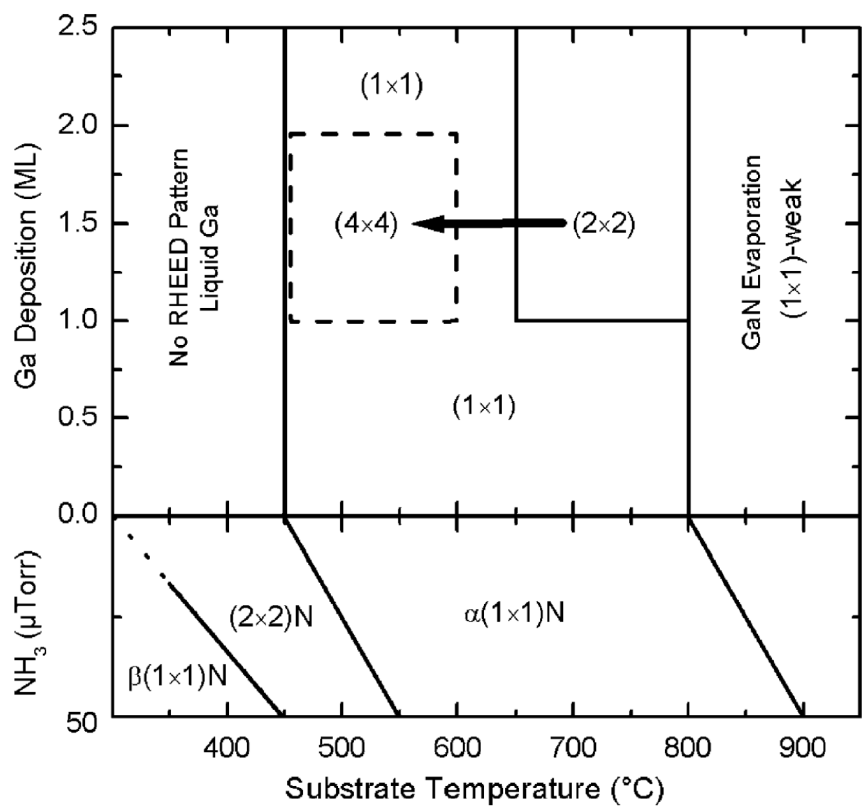

Fig. 2

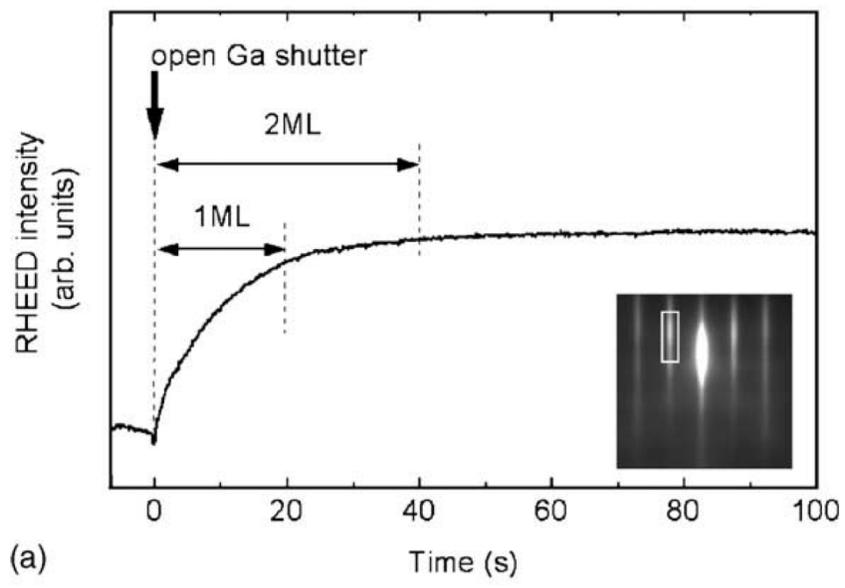
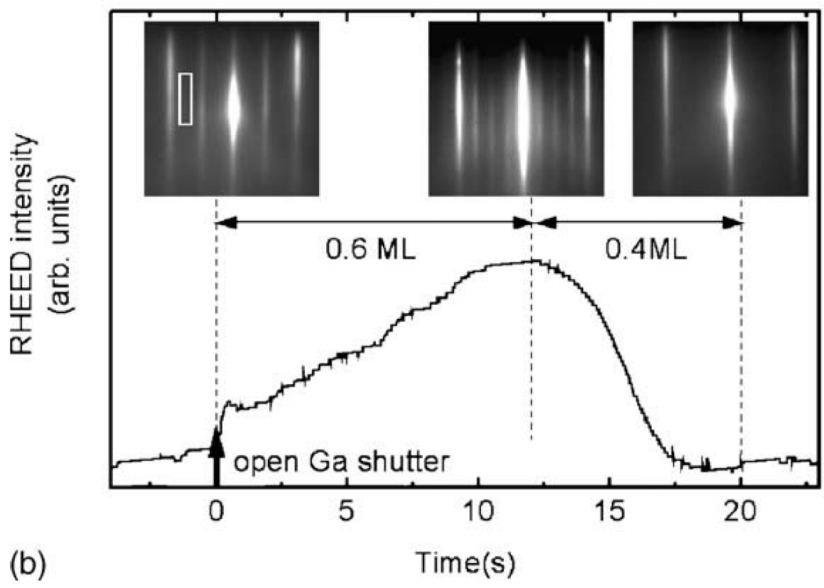

Fig. 3

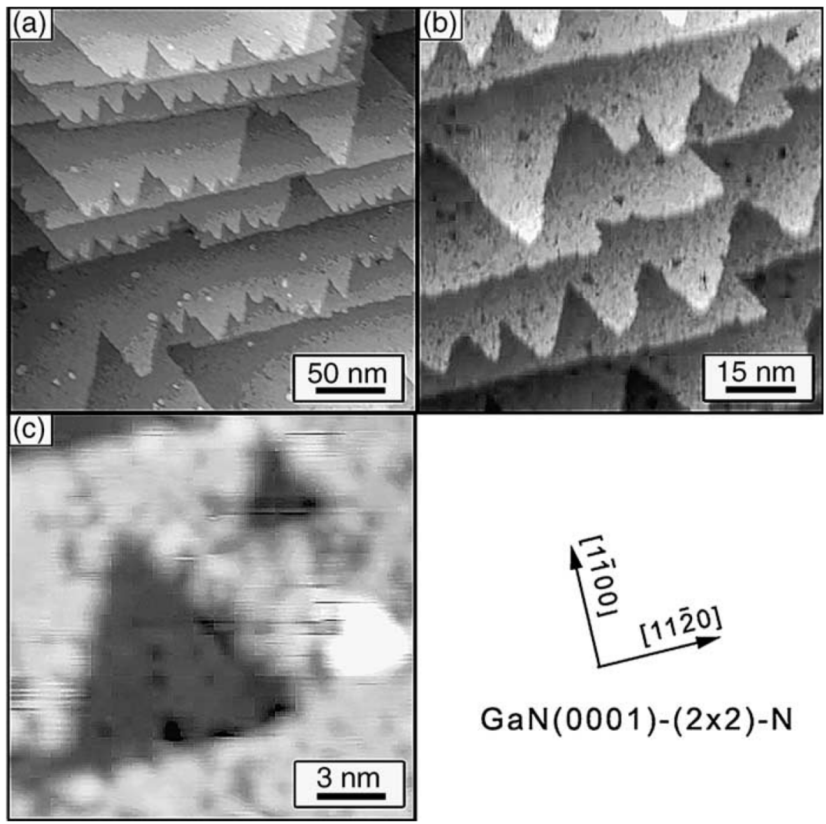

Fig. 4

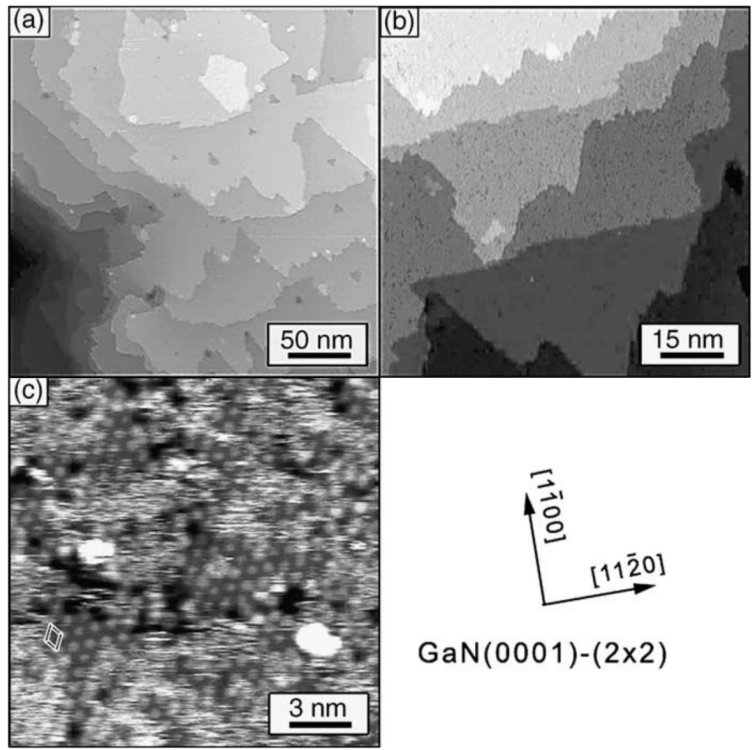

Fig. 5

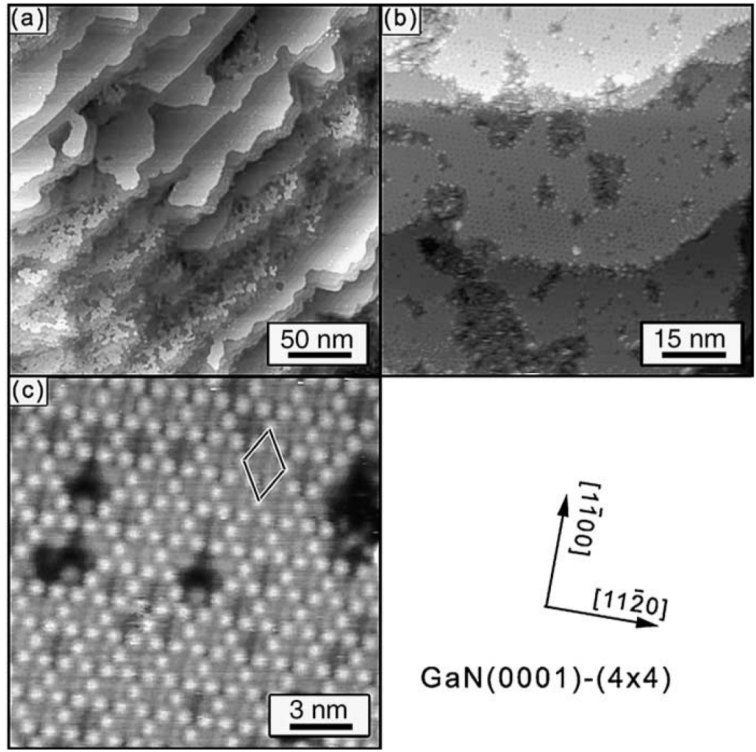

Fig. 6

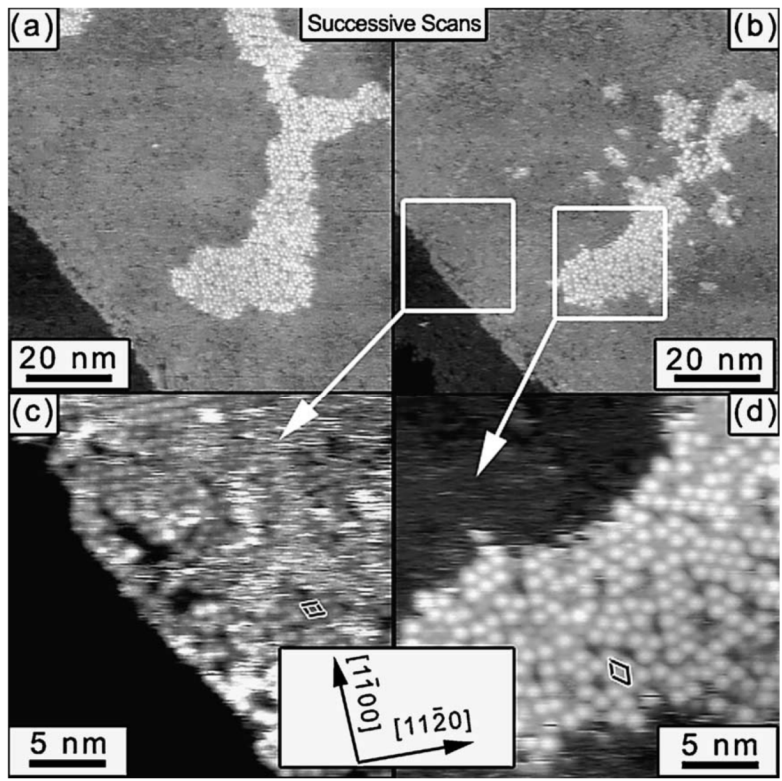

Fig. 7